\documentclass[final,5p,times,twocolumn]{elsarticle}

\usepackage{graphicx}
\usepackage{subfig}
\usepackage{amssymb}
\usepackage{amsmath}

\usepackage{lineno}

\usepackage{siunitx}				
\usepackage[]{tikz}
\usepackage{pgfplots}
\pgfplotsset{compat =1.9}

\usepackage{caption} 
\usepackage[labelsep=period]{caption}
\usepackage[labelfont=bf]{caption}

\usepackage{float}
\usepackage{stfloats}
\usepackage{color}
\journal{Acta Materialia}

\begin{document}

\begin{frontmatter}

\title{Crystallographic relationships of T-/S-phase aggregates in an Al--Cu--Mg--Ag alloy}

\author[NTNU]{Jonas K. Sunde\corref{cor1}}
\ead{jonas.k.sunde@ntnu.no}
\author[Cambridge]{Duncan N. Johnstone}
\author[SINTEF]{Sigurd Wenner}
\author[NTNU]{Antonius T.J. van~Helvoort} 
\author[Cambridge]{Paul A. Midgley}
\author[NTNU]{Randi Holmestad}
\address[NTNU]{Department of Physics, Norwegian University of Science and Technology (NTNU), H\o gskoleringen 5, N-7491 Trondheim, Norway}
\address[Cambridge]{Department of Materials Science and Metallurgy, University of Cambridge, 27 Charles Babbage Road, CB3 0FS Cambridge, UK}
\address[SINTEF]{Materials and Nanotechnology, SINTEF Industry, H\o gskoleringen 5, N-7491 Trondheim, Norway}
\cortext[cor1]{Corresponding author}
\fntext[fn1]{https://doi.org/10.1016/j.actamat.2018.12.036}
\fntext[fn2]{$\textcopyright$ 2018. This manuscript version is made available under the CC-BY-NC-ND 4.0 license http://creativecommons.org/licenses/by-nc-nd/4.0/}

\begin{abstract}
T-(Al$_{20}$Cu$_{2}$Mn$_{3}$) phase dispersoids are important for limiting recovery and controlling grain growth in Al-Cu alloys. However, these dispersoids can also reduce precipitation hardening by acting as heterogeneous nucleation sites and may lead to increased susceptibility towards pitting corrosion when galvanically coupled with S-(Al$_2$CuMg) phase precipitates. The interplay between T- and S-phases is therefore important for understanding their effect on the mechanical and electrochemical properties of Al-Cu-Mg alloys. Here, the crystallographic relationships between the T-phase, S-phase, and surrounding Al matrix were investigated in an Al-1.31Cu-1.14Mg-0.13Ag-0.10Fe-0.28Mn (at.\%) alloy by combining scanning precession electron diffraction with misorientation analysis in 3-dimensional axis-angle space and correlated high-resolution transmission electron microscopy. Orientation relationships are identified between all three phases, revealing S-T orientation relationships for the first time. Differences in S-Al orientation relationships for precipitates formed at T-phase interfaces compared to their non-interfacial counterparts were also identified. These insights provide a comprehensive assessment of the crystallographic relationships in T-/S-phase aggregates, which may guide future alloy design. 
\end{abstract}


\end{frontmatter}

\section{Introduction}

2xxx series Al alloys are Cu-containing age-hardenable alloys widely used in the aerospace industry due to their high strength-to-weight ratio, good formability, and high damage tolerance \cite{Starke, Totten}. Alloys based on the Al-Cu-Mg system are particularly common owing to their high fracture toughness and fatigue resistance \cite{Williams, Dursun}. Al-Cu-Mg alloys obtain most of their strength from a distribution of atomic clusters, Guinier–-Preston-–Bagaryatsky (GPB) zones \cite{Kovarik1}, and precipitates of different metastable phases that are formed throughout the Al matrix during heat treatment. Commercial alloys typically have an atomic ratio Cu:Mg$>$1 which 
eventually leads to formation of equilibrium S-(Al$_{\text{2}}$CuMg) and/or $\theta$-(Al$_{\text{2}}$Cu) precipitates preceded by their precursor phases \cite{Guinier, Wang, Ringer2, Wang3}. Ag additions to this alloy system have been found to increase strength by modifying and enhancing the age hardening response \cite{Vietz, Taylor, Bakavos}. This is achieved by promoting $\Omega$-(Al$_{\text{2}}$Cu) phase formation over $\theta$'(Al$_{\text{2}}$Cu) \cite{Lumley, Sano} leading to a denser distribution of finer precipitates. Mn additions reduce the detrimental effects on mechanical properties from Fe impurities by forming dispersoids during high temperature homogenisation. 
In Al-Cu-Mg alloys the main dispersoid phase is T (Al$_{20}$Cu$_{2}$Mn$_{3}$, Bbmm, \textit{a} = 23.98 \AA, \textit{b} = 12.54 \AA, \textit{c} = 7.66 \AA\ \cite{Shen}) which effectively pins grain boundaries, limits recrystallisation, accumulates dislocations, and resists recovery after forming \cite{FengWX, Castillo, Cheng}. 
The T-phase dispersoid can increase the sensitivity to micro-crack initiation but can on the other hand prevent fast and continuous crack propagation \cite{Feng2}. The T-Al interface can act as a heterogeneous nucleation site for phases such as $\Omega$, $\theta$($\theta'$), and S(S$'$) \cite{Feng,Mukhopadhyay,Wang2} creating dispersoid aggregates and reducing normal intra-granular precipitation.

Although the contribution of Cu and Mg in precipitation of various phases results in a higher strength and mechanical performance, the phases formed generally lead to a significant drop in corrosion resistance \cite{Dursun}. 
Extensive research has been devoted to study different aspects of corrosion in these alloys, including localized corrosion, galvanic coupling between phases and stress-corrosion cracking \cite{Urushino, Parvizi, Parvizi2, Wang_corr2, Birbilis}. Several studies point out the S- and T-phase as most important particularly with regards to initiation of pitting corrosion, especially when the two phases are in contact forming a galvanic couple \cite{Wang_corr}. 

\begin{figure}[h]
\centering\includegraphics[width=\linewidth]{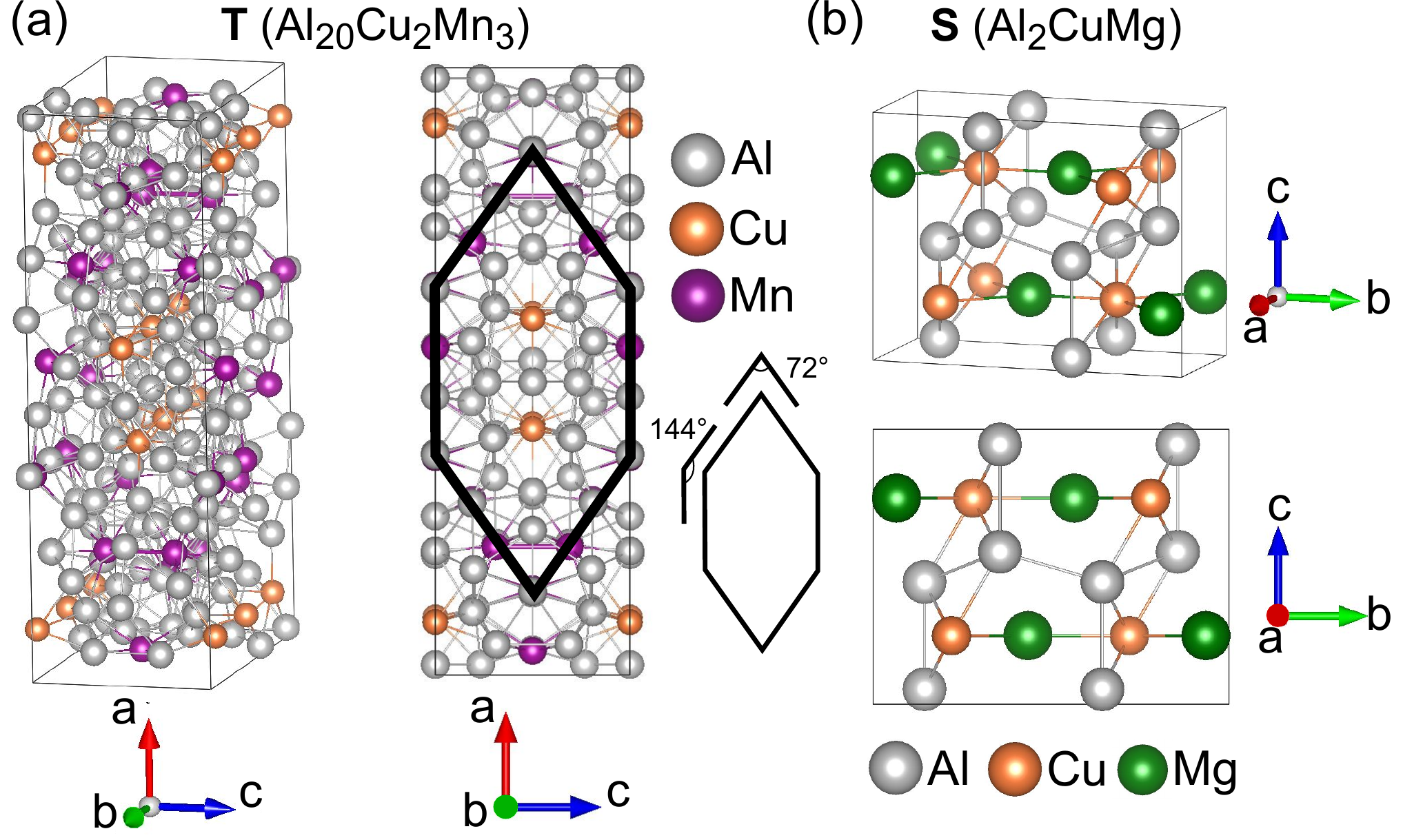}
\caption{Crystal structures of the (a) T- and (b) S-phase. A flattened hexagonal subunit in the T-phase is indicated and is relevant in twinning of this structure.}
\label{fig:structures}
\end{figure}

T-phase dispersoids tend to adopt rod-like morphologies with their long axes along $\langle001\rangle_\text{Al}$ directions. In cross-section, the morphology is typically a lath or a shell-shaped structure \cite{Chen3}. Internal faulting, particularly twinning, is common, and successive twinning transforms the cross-section from a lath to a shell-shaped structure \cite{Chen_new}. 
The crystal structure of the S-phase has been subject to much debate but there is now broad consensus around the Perlitz-Westgren structure (Al$_2$CuMg, Cmcm, \textit{a} = 4.00 \AA, \textit{b} = 9.23 \AA, \textit{c} = 7.14 \AA) \cite{Perlitz, Wolverton}. The crystal structures of the T- and S-phase are shown in \textbf{Fig. \ref{fig:structures}}.
\\
\\

Crystallographic orientation relationships (ORs) between the T-phase, S-phase, and Al matrix, as well as the internal structure of the T-phase, have been been the subject of numerous studies across a range of compositions and thermo-mechanical treatments. Reported ORs are summarised in \textbf{Table \ref{tab:TAl_ORs}} and \textbf{Table \ref{tab:SAl_ORs}}.
Three ORs have been observed between the T-phase and Al matrix \cite{Chen2}, all with the [010]$_\text{T}$ axis aligned along 
$\langle001\rangle_\text{Al}$ but with different coincident planes. 
When viewed along the $[010]_\text{T}$ axis, the twinning is seen as a $\sim$$36^{\circ}$ rotation about the axis with \{101\}$_\text{T}$ as the twin plane, which can yield structures exhibiting pseudo 10-fold symmetry \cite{Feng3}.

Different crystallographic relationships between the S-phase and Al matrix have been reported, the most frequent of which is that discovered by Bagaryatsky \cite{Bagaryatsky} (OR(I) in \textbf{Table \ref{tab:SAl_ORs}}) corresponding to type I S-phase. 
However, rotations of several degrees away from this S-Al OR are observed. This has led several authors \cite{Wang3, Majimel2, Winkelman, Kovarik, Radmilovic} to distinguish a second OR (e.g. OR(II-IV) in \textbf{Table \ref{tab:SAl_ORs}}) rotated by 3$^{\circ}$ to as much as 9$^{\circ}$ about the [100]$_{\text{S}}$//[001]$_\text{Al}$ axis away from OR(I), and is potentially associated with a second type of S-phase (type II). It has been suggested that the two ORs may be extrema of a continuous or near continuous distribution of rotation angles \cite{Winkelman}. The rotation has been rationalized in terms of a competition between elastic strains due to lattice mismatch between coherent interfaces formed at each OR \cite{Kovarik}. Some studies found that type II S-phase grows at the expense of type I, and that type II is the more stable phase \cite{Wang3, Styles}. Studies conducted by Styles \textit{et al.} \cite{Styles, Styles2} concluded that there are no significant differences in crystal structure between the two types, but that type I has a deficiency of Cu which may explain observed variations in lattice parameters between the two types. 
S-phases adopting OR(I) tend to be lath-like with atomically sharp interfaces, whereas precipitates following the second OR tend to be rod-shaped with stepped interfaces \cite{Radmilovic}. The ratio of type I relative to type II S-phases is among others dependent on ageing times and temperature \cite{Styles}, as well as quenching rate from homogenisation and whether cold work is applied prior to ageing \cite{Wang3, Parel}. S-T ORs and S-Al ORs for S-phase precipitates nucleated on T-phase dispersoids have been less studied \cite{Feng}.
\\

In this work, scanning precession electron diffraction (SPED) is applied in combination with misorientation analysis in axis-angle space and high-resolution imaging to understand the structure of dispersoid aggregates comprised of T-phase dispersoids and S-phase precipitates surrounded by the Al matrix. In addition, a comparison is made between S-Al misorientations for S-phase precipitates decorating T-phase dispersoids and those located away from T-phase interfaces. Knowledge of these crystallographic relationships will enhance understanding of the effect of T-/S-phase aggregates on mechanical and electrochemical properties in Al-Cu-Mg alloys.

\begin{table}[h]
\centering
\caption {Reported T-Al orientation relationships \cite{Chen2}. $\mathbf{n}$ denotes a unit vector that runs parallel to the axis of rotation.}
\label{tab:TAl_ORs}
\begin{tabular}{ c | c | c }
 OR & Parallelism & Axis-angle \\ 
\hline
(I) & $\lbrace200\rbrace_{\text{T}}$ // $\lbrace200\rbrace_{\text{Al}}$, $\langle010\rangle_{\text{T}}$ // $\langle001\rangle_{\text{Al}}$ (// $\mathbf{n}$) & $\mathbf{n}, 0^{\circ}$  \\
(II) & $\lbrace200\rbrace_{\text{T}}$ // $\lbrace40\overline{3}\rbrace_{\text{Al}}$, $\langle010\rangle_{\text{T}}$ // $\langle001\rangle_{\text{Al}}$ & $\mathbf{n}, 36.87^{\circ}$\\
(III) & $\lbrace200\rbrace_{\text{T}}$ // $\lbrace301\rbrace_{\text{Al}}$, $\langle010\rangle_{\text{T}}$ // $\langle001\rangle_{\text{Al}}$ & $\mathbf{n}, 18.43^{\circ}$
\end{tabular}
\end{table}

\begin{table}[h]
\centering
\caption {Reported S-Al orientation relationships I \cite{Bagaryatsky}, II \cite{Majimel2}, III \cite{Winkelman, Kovarik}, and IV \cite{Radmilovic}.}
\label{tab:SAl_ORs}
\begin{tabular}{ c | c | c }
 OR & Parallelism & Axis-angle \\
\hline
(I) & $\lbrace001\rbrace_{\text{S}}$ // $\lbrace012\rbrace_{\text{Al}}$, $\langle100\rangle_{\text{S}}$ // $\langle001\rangle_{\text{Al}}$ (// $\mathbf{n}$) & $\mathbf{n}, 26.57^{\circ}$\\
(II) & $\lbrace001\rbrace_{\text{S}}$ // $\lbrace0\overline{5}2\rbrace_{\text{Al}}$, $\langle100\rangle_{\text{S}}$ // $\langle001\rangle_{\text{Al}}$ & $\mathbf{n}, 21.80^{\circ}$\\
(III) & $\lbrace0\overline{2}1\rbrace_{\text{S}}$ // $\lbrace014\rbrace_{\text{Al}}$, $\langle100\rangle_{\text{S}}$ // $\langle001\rangle_{\text{Al}}$ & $\mathbf{n}, 18.84^{\circ}$\\
(IV) & $\lbrace043\rbrace_{\text{S}}$ // $\lbrace021\rbrace_{\text{Al}}$, $\langle100\rangle_{\text{S}}$ // $\langle001\rangle_{\text{Al}}$ & $\mathbf{n}, 17.55^{\circ}$
\end{tabular}
\end{table}

\section{Material \& Methods}

\subsection{Material}

The nominal composition of the Al alloy studied in this work is shown in \textbf{Table \ref{tab:composition}}. The as-received material was an extruded rod pre-heated to $\SI{400}{\degreeCelsius}$ and extruded at $\SI{390}{\degreeCelsius}$. From the extruded rod a cylinder (\O\ = $\SI{20}{\milli\meter}$, height = $\SI{10}{\milli\meter}$) was cut and solution heat treated at $\SI{440}{\degreeCelsius}$ for $\SI{1}{\hour}$ before water-quenched to room temperature. The material was then directly set to artificial ageing at $\SI{170}{\degreeCelsius}$ conducted in an oil-bath, avoiding any natural ageing effects. The material was studied in an over-aged condition, which was obtained after $4$ days of ageing.

\begin{table}[h]
\centering
\caption {Nominal composition of the Al alloy studied.}
\label{tab:composition}
\begin{tabular}{l|cccccc}
Element & Al   & Cu & Mg & Ag  & Fe & Mn \\ \hline
at.\% & bal. & 1.31 & 1.14 & 0.13 & 0.10 & 0.28 \\
wt.\% & bal. & 3.00 & 1.00 & 0.50 & 0.20 & 0.55 
\end{tabular}
\end{table}

Electron transparent thin film specimens were prepared from $\SI{3}{\milli\meter}$ diameter discs of material ground to a thickness of $\sim$$\SI{100}{\micro\meter}$ before further thinning by electrolytic polishing. Electro-polishing was performed using a Struers Tenupol-5 operated at a voltage of $\SI{20}{\volt}$ (current $\SI{0.2}{\ampere}$). The electrolytic solution comprised a 2:1 mixture of methanol:nitric acid and was held at a temperature in the range $-\SI{30}{\degreeCelsius}$ to $-\SI{25}{\degreeCelsius}$. Prior to SPED and high resolution microscopy, the specimens were cleaned \mbox{using} a \mbox{Fischione} 1020 Plasma Cleaner to reduce the risk of carbon \mbox{contamination} build-up during data acquisition.

\subsection{Electron microscopy}

High-resolution transmission electron microscopy (HRTEM) and SPED were performed using a JEOL 2100F (S)TEM operated at 200 kV and fitted with a NanoMEGAS ASTAR system \cite{Astar, Moeck} to enable the simultaneous scan and acquisition of precession electron diffraction patterns at each probe position. SPED was performed with the microscope operated in nanobeam diffraction mode. The probe convergence semi-angle was measured as $\SI{1.0}{\milli\radian}$. The precession angle employed was $\SI{0.5}-\SI{1.0}{\degree}$ and the precession frequency was set to 100 Hz. The scan step size was in the range $\SI{0.76}-\SI{2.28}{\nano\meter}$ and the exposure time per pixel was $\SI{20}-\SI{40}{\milli\second}$. Diffraction patterns were recorded using a Stingray camera photographing the microscope's fluorescent screen. The double-rocking probe required for PED was aligned following the method detailed by Barnard \textit{et al.} \cite{Barnard}. High-angle annular dark-field STEM (HAADF-STEM) was performed using a double corrected JEOL ARM 200F microscope operated at $\SI{200}{\kilo\volt}$ using a \mbox{detector} collection angle of \SIrange{42}{178}{\milli\radian}.

\subsection{Phase \& orientation mapping}

Phase and orientation maps were formed using the pattern matching approach of Rauch \textit{et al.} \cite{Rauch} in which the 2-dimensional PED pattern recorded at each probe position in a 2-dimensional area scan is matched against a library of simulated diffraction patterns, for all asymmetric orientations of the expected phases. Prior to this pattern matching, a background subtraction was applied to each PED pattern using a routine implemented in the pyXem Python library \cite{pyxem,Pena,Pena2}, and the template matching parameters in the ASTAR software were tuned to obtain good agreement between the matching results and HRTEM images of the same particles, as detailed in \textbf{Supplementary Information (SI)}.

Orientation mapping results were analysed using the Matlab toolbox MTEX \cite{Bachmann}, following procedures described by Krakow \textit{et al.} \cite{Krakow2}. Crystallographic domains in the dispersoid aggregates were typically defined using regions with a common phase and orientation within a threshold of 10$^{\circ}$, which was found to give good agreement with the same phases observed in HRTEM images. Orientation relationships between these crystallographic domains were investigated by calculating the misorientation between neighbouring pixels across all domain boundaries. This misorientation data was then visualised, for each type of phase boundary by plotting the disorientation between adjacent domains as a vector in the appropriate symmetry reduced region (\textit{fundamental zone}) of a 3-dimensional misorientation space \cite{Krakow2}. The vector space chosen was the \textit{axis-angle space} in which a disorientation is represented by a vector $\boldsymbol{\rho}$:

\begin{equation}
\boldsymbol{\rho} = \omega \mathbf{n}
\label{eq:rho}
\end{equation}

where $\mathbf{n}$ is a unit vector parallel to the axis of rotation and $\omega$ is the angle of rotation [$^{\circ}$]. This representation is preferred over other mappings because the rotation angle is simply read from the plot and it is sufficient for visualisation.

\section{Results}

\begin{figure}[h!]
\centering\includegraphics[width=.85\linewidth]{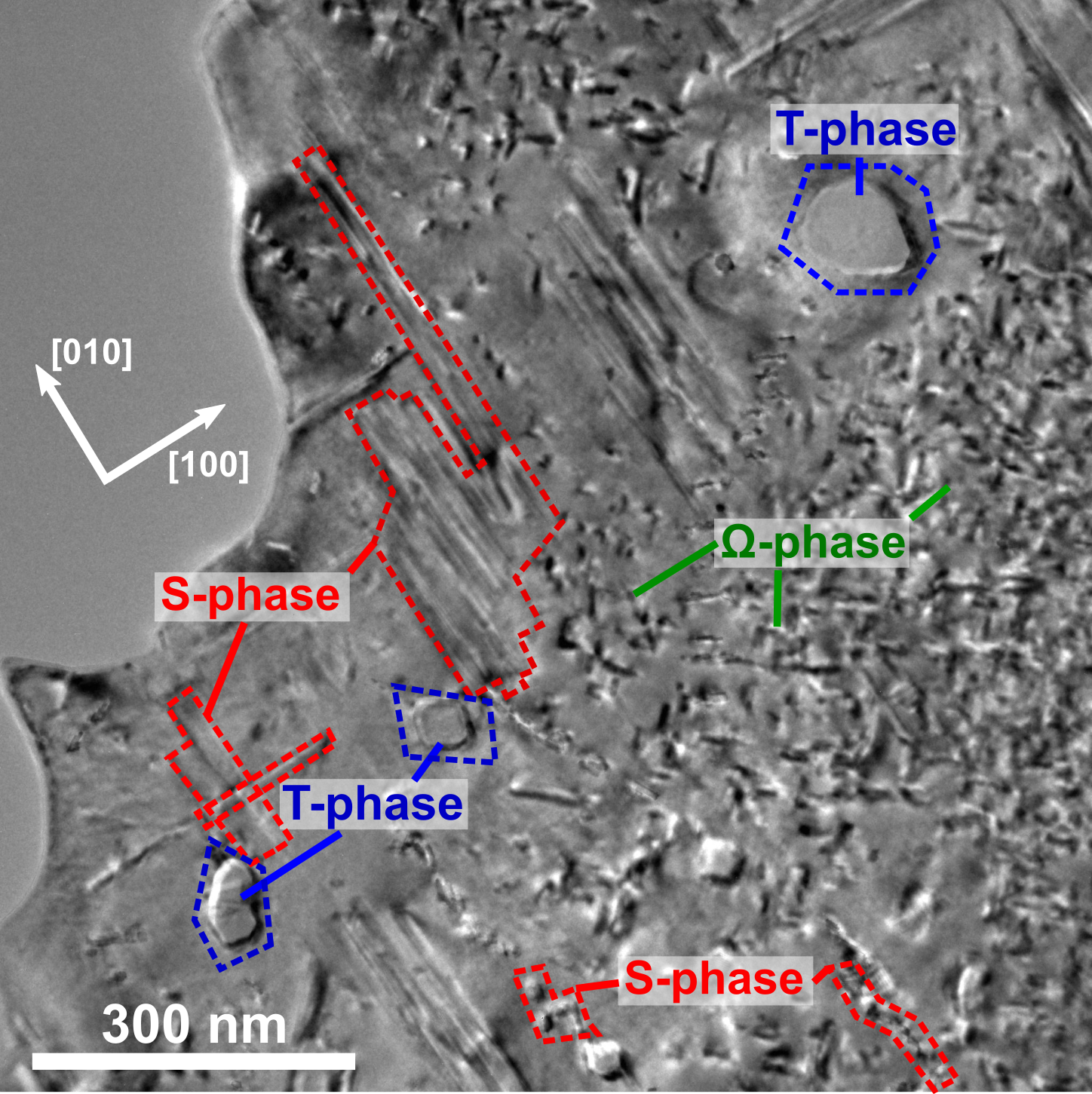}
\caption{TEM image of the over-aged Al-Cu-Mg-Ag alloy microstructure as viewed near the [001]$_\text{Al}$ zone axis. The main phases observed are indicated.}
\label{fig:tem}
\end{figure}

\begin{figure}[h!]
\centering\includegraphics[width=.85\linewidth]{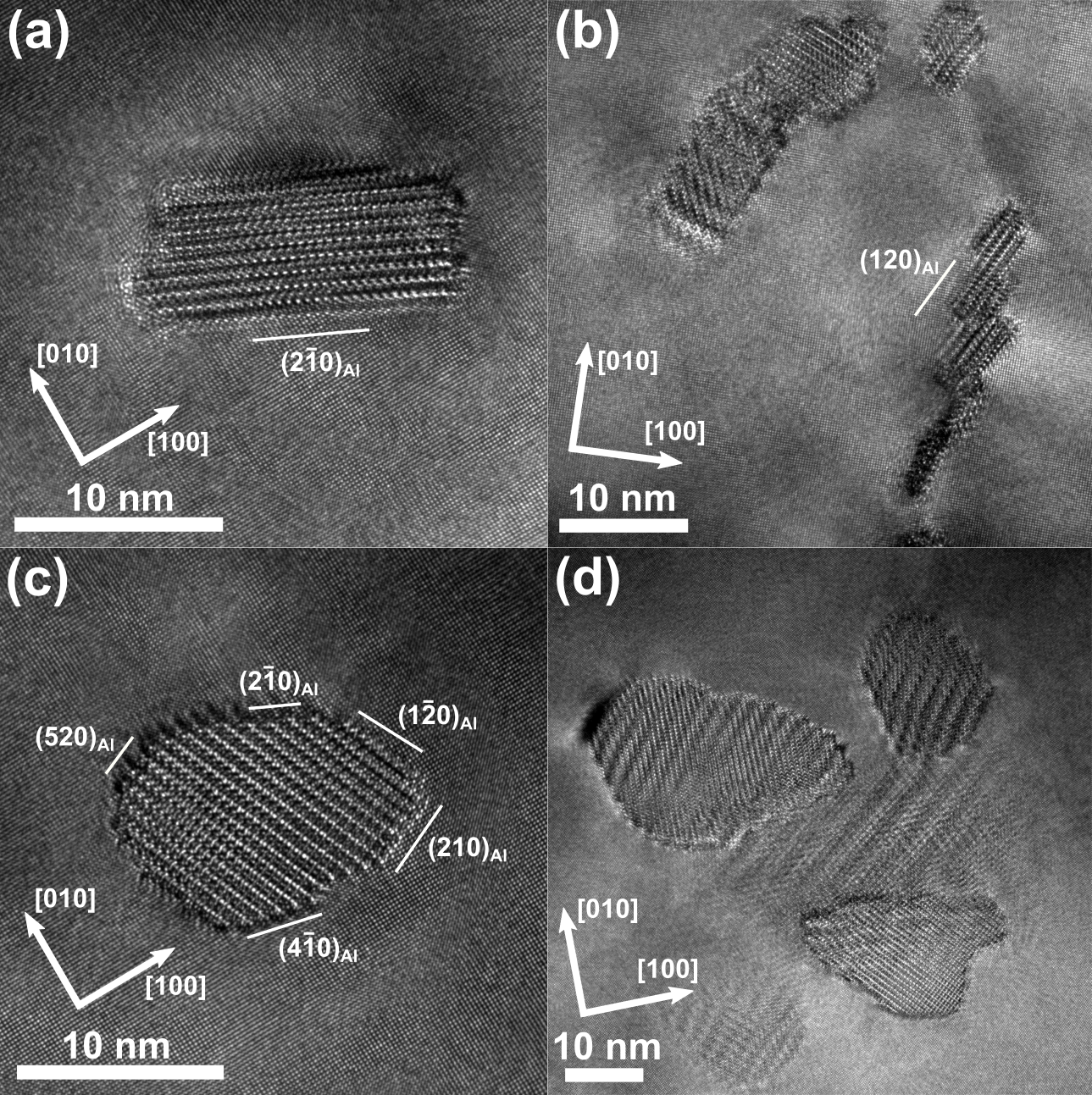}
\caption{Different variations of S-phases in the alloy microstructure as observed near the [001]$_\text{Al}$ zone axis. (a) A single lath-shaped S-phase. (b) A wall-structure of adjoining S-phases nucleated on a dislocation network. (c) A single rod-shaped S-phase. (d) A cluster of coarsened S-phases.}
\label{fig:S-phases}
\end{figure}

\subsection{TEM observations}

\begin{figure*}[b] 
\centering\includegraphics[width=1.\linewidth]{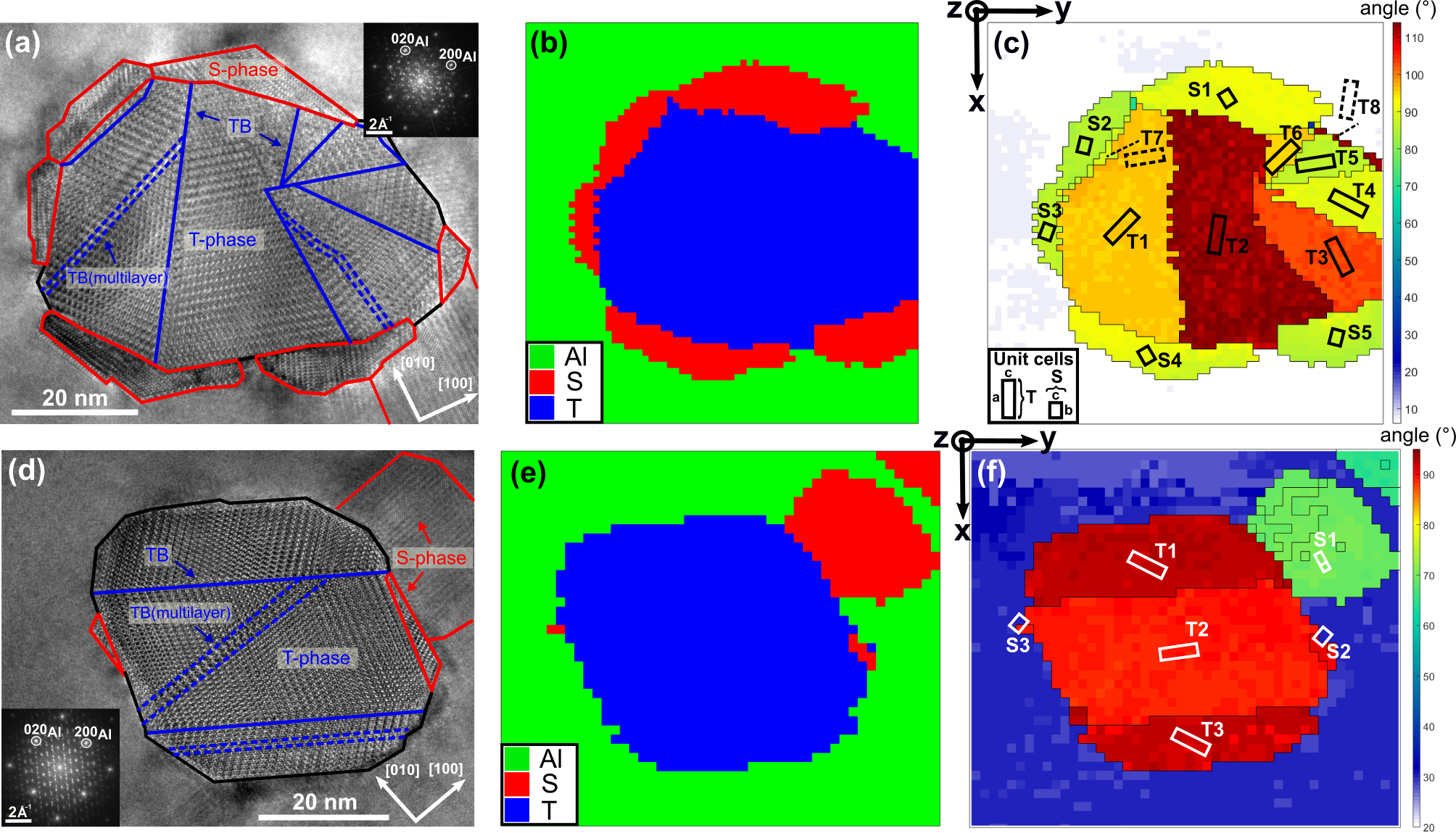}
\caption{Structure of a (a-c) shell-shaped and (d-f) lath-shaped T-/S-phase aggregate. (a,d) HRTEM images of the aggregates with S-phases and T-phase TBs/TB multilayers indicated. Inserts show the image fast fourier transforms. (b,e) Phase maps obtained via template matching of SPED data. (c,f) Orientation maps showing the disorientation angle $\omega$ taken about an axis $\boldsymbol{\rho}$ (\textbf{Eq. \ref{eq:rho}}) at each probe position relative to the specimen reference frame (x,y,z). The average orientation of each labelled domain/phase is indicated by drawn unit cells of the T- and S-phase (not to scale).}
\label{fig:mapping}
\end{figure*}

Conventional TEM imaging was performed to obtain an overview of the phases present, which were identified based on characteristic morphology and lattice structure in high-resolution images, as shown in \textbf{Fig. \ref{fig:tem}}, \textbf{Fig. \ref{fig:S-phases}}, and \textbf{Fig. \ref{fig:mapping}}(a,d). A high number density of $\Omega$-phase precipitates were observed throughout the Al matrix, as well as elongated S-phase precipitates and T-phase dispersoids. The $\Omega$-phase exists as thin, hexagonal shaped plates formed on \{111\}$_\text{Al}$. S-phase precipitates were more inhomogeneously distributed, growing as rods or laths on \{021\}$_\text{Al}$ planes and extending along $\langle$100$\rangle_\text{Al}$ directions. S-phase precipitates were frequently observed clustered together, having formed heterogeneously on T-phase dispersoids or in wall-structures of adjoining S-phase precipitates on underlying dislocation networks, in agreement with previous reports \cite{Feng, Feng4} (see \textbf{Fig. \ref{fig:S-phases}}(b,d) and \textbf{Fig. \ref{fig:mapping}}(a,d)). Both lath- and rod-shaped S-phase cross-sections were observed, laths being more frequent. A smaller number of individual S-phase precipitates, likely formed by homogeneous bulk nucleation, were also observed (\textbf{Fig. \ref{fig:S-phases}}(a,c)). Moir\'e fringes can be observed in some images, e.g. \textbf{Fig. \ref{fig:S-phases}}(b,d), which are attributed primarily to overlap between the S-phase cross-sections and surrounding Al.  Whilst this suggests a more diverse range of precipitate morphologies than is evident from these projection images this does not affect assessment of orientation relationships between phases.

Dispersoids were identified as T-phase and all of these were decorated by S-phase precipitates at prior T-Al interfaces (see \textbf{Fig. S1-24} in \textbf{SI}). Typically, \SIrange{1}{10}{} S-phases were observed at each dispersoid. 
No $\Omega$-phases were observed at the T-phase interfaces. Morphologically, both shell-shaped (\textbf{Fig. \ref{fig:mapping}}(a)) and lath-shaped (\textbf{Fig. \ref{fig:mapping}}(d)) T-phase dispersoids were observed, and all exhibited some degree of rotation-twinned substructure. T-phase cross-section diameters or diagonals were measured in the range \SIrange{45}{75}{\nano\meter}, and cross-section aspect ratios varied between $1.1$-$1.9$.
S-phase precipitates were observed to grow both parallel and perpendicular to the T-phase elongation axis. S-phase precipitate lengths were measured as $\SI{136(19)}{\nano\meter}$ and $\SI{146(24)}{\nano\meter}$ for S-phases at T-phase interfaces (i.e. S-phases with elongation perpendicular to the T-phase axis) and non-interfacial S-phases, respectively. Cross-section areas for the two categories were $\SI{93(8)}{\nano\meter^2}$ and $\SI{69(8)}{\nano\meter^2}$, respectively. This indicates a slight coarsening of S-phase precipitates decorating T-phase dispersoids.

\subsection{Dispersoid aggregates}

Dispersoid aggregates comprise the T-phase dispersoid, interfacial S-phases, and surrounding Al matrix. The structure of 10 such aggregates was investigated in detail using correlated HRTEM and SPED data, as shown in \textbf{Fig. S1-20}. Two examples are presented in \textbf{Fig. \ref{fig:mapping}}, showing a shell- and lath-shaped T-phase exhibiting pronounced and limited rotation-twinning, respectively. 
Comparing HRTEM images and SPED mapping results from the same dispersoid aggregates demonstrates that the primary crystallographic features were accurately captured in the SPED data analysis. T-phase substructures such as anti-phase boundaries (APBs), micro twins, and twin boundaries (TBs) confined to a limited number of hexagon subunits in width (multilayers) \cite{Feng3, Wang__} could not be resolved by SPED, but are often associated with pixels of lower reliability and/or index value in pattern matching (\textbf{Fig. S1-20}). In total, 43 S-phases at the interface of T-phases were mapped by SPED. The disorientation data for each type of phase boundary shown in the following is the combined data extracted from all SPED scans recorded in the present work.

\subsubsection{T-phase orientation relationships}

\begin{figure}[h]
\centering\includegraphics[width=0.8\linewidth]{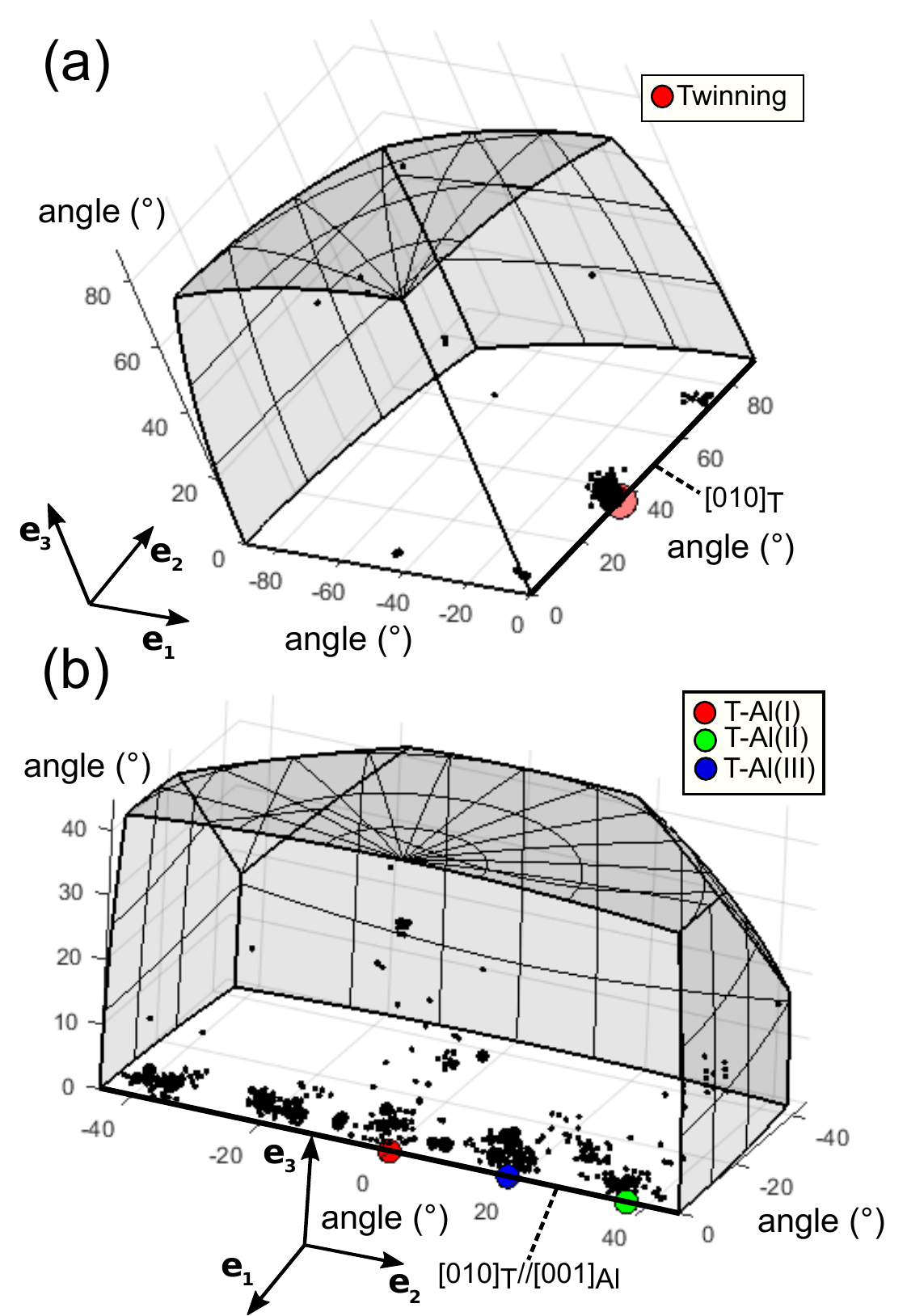}
\caption{T-phase disorientations and previously reported orientation relationships plotted in corresponding fundamental zones of axis-angle space. (a) T-T disorientations across 34 boundaries. (b) T-Al disorientations across 51 boundaries.}
\label{fig:t-disori}
\end{figure}

Crystallographic relationships between rotated domains of the T-phase (T-T) and across the T-Al interface were assessed by plotting disorientations within the corresponding fundamental zones of axis-angle space as shown in\textbf{ Fig. \ref{fig:t-disori}}, with previously reported T-Al ORs (\textbf{Table \ref{tab:TAl_ORs}}) highlighted. T-T domain boundary disorientations (\textbf{Fig. \ref{fig:t-disori}}(a)) are clustered near a $\sim$36$^{\circ}$ rotation about $[010]_{\text{T}}$ corresponding to twinning \cite{Feng3}. Two smaller clusters are also observed at 0$^{\circ}$ and $\sim$72$^{\circ}$, the latter corresponding to twice the twinning angle. Both are associated with disorientations across domains in the rotation-twin centres, i.e. the regions from which the twin domains seemingly emanate. In this region, each domain shares a small boundary to other domains that can be rotated relative to it by 0$^{\circ}$, $\sim$36$^{\circ}$, or $\sim$72$^{\circ}$. 
Data points away from these disorientation clusters were likely misindexed pixels as a result of local disorder, e.g. the unresolved APBs, micro twins or TB multilayers. 
The spread in T-T TB disorientations (FWHM of angle distribution) of $\sim$1.2$^{\circ}$ is taken as an indication of angular resolution, which is consistent with estimated angular resolution for spot pattern based indexation \cite{Rauch2}.

T-Al disorientations (\textbf{Fig. \ref{fig:t-disori}}(b)) form 3 main clusters near previously reported T-Al ORs. The 2 clusters situated at negative angles along the [010]$_{\text{T}}$//[001]$_\text{Al}$ axis correspond, by symmetry, with those at equivalent positive values, i.e. near ORs (II) and (III). There is a spread of rotations $\sim$$ 4^{\circ}$ away from the exact angles described by the ORs, primarily about the [010]$_{\text{T}}$//[001]$_\text{Al}$ axis since most data points are distributed parallel to this axis. This implies that there exists a deviation of $\pm 4^{\circ}$ from exact OR parallelisms across the T-Al interface. 
The cluster spread is less pronounced perpendicular to the [010]$_{\text{T}}$//[001]$_\text{Al}$ axis, implying that the [010]$_{\text{T}}$ elongation axis remains reasonably parallel to the [001]$_\text{Al}$ direction. A 5$^{\circ}$ radius sphere of disorientations about OR(I), OR(II), and OR(III) accounts for 7\%, 30\%, and 50\% of all disorientation data points, respectively.

\begin{figure}[h]
\centering\includegraphics[width=.95\linewidth]{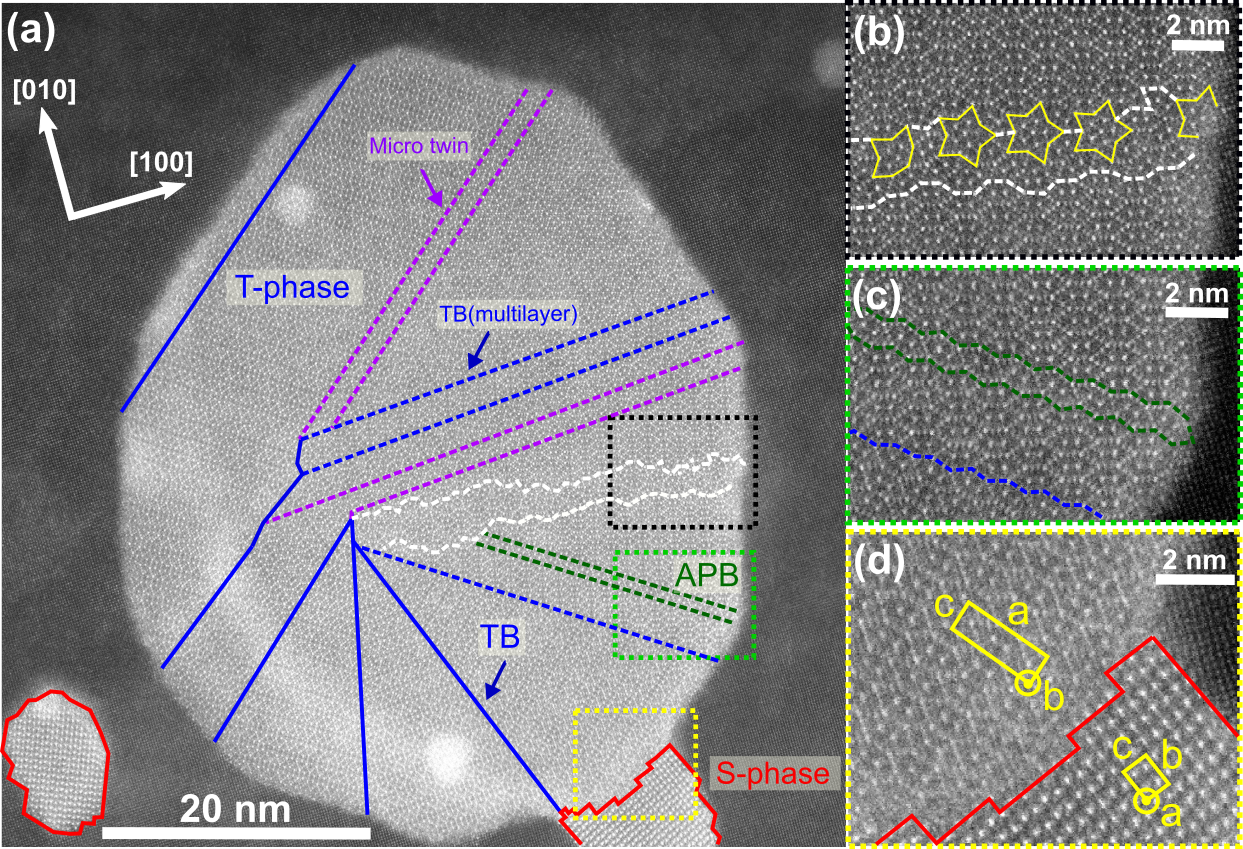}
\caption{(a) HAADF-STEM image of a T-phase dispersoid with a S-phase precipitate at the prior T-Al interface as viewed near the [001]$_\text{Al}$ zone axis. TBs/TB multilayers, micro twins, APB, and a transition region are observed. (b) Transition region comprising different tessellations of the hexagonal subunit. (c) An APB and a TB multilayer. (d) S-T phase boundary with unit cells indicated.}
\label{fig:hrstem}
\end{figure}

The substructure of T-phase dispersoids are observed in HAADF-STEM, as shown in \textbf{Fig. \ref{fig:hrstem}}. APBs (\textbf{Fig. \ref{fig:hrstem}}(c)) form where there exists a band one single hexagon subunit in width rotated (by the twinning angle) with respect to the surrounding twin domain. A micro twin forms when a second band rotated with respect to the first appears. 
A TB multilayer appears as a narrow band, usually \SIrange{2}{5}{} hexagon subunits in width. 
Another more complex feature is transition regions (\textbf{Fig. \ref{fig:hrstem}}(b)), which comprise different geometrical structures formed by various hexagon subunit tessellations. 
The T-phase substructures that extend to the T-Al interface are likely the main explanation for misindexed data points in \textbf{Fig. \ref{fig:t-disori}}(b) that are spread at significant distances away from the main observed disorientation clusters. 
The bright regions in \textbf{Fig. \ref{fig:hrstem}} are primarily due to incorporation of Ag in the S- and T-phase. Additional HAADF-STEM images are presented in \textbf{Fig. S23} and \textbf{Fig. S24}.

\subsubsection{S-Al orientation relationships}

\begin{figure}[h]
\centering\includegraphics[width=0.7\linewidth]{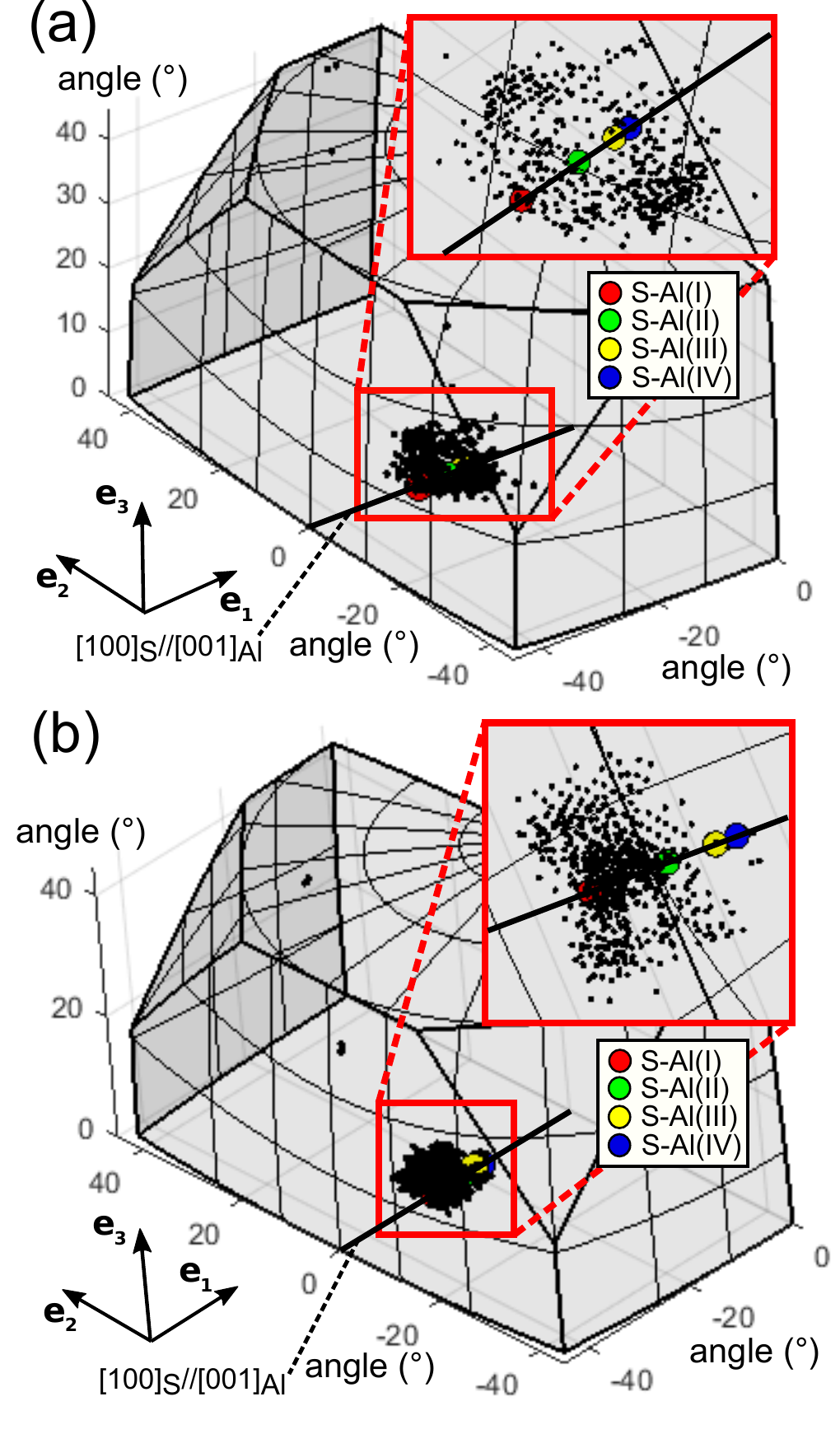}
\caption{S-Al disorientations and previously reported orientation relationships plotted in fundamental zones of axis-angle space. (a) S-Al disorientations across 43 boundaries for S-phases situated at T-phase interfaces. (b) S-Al disorientations across 57 boundaries for S-phases placed away from T-phase interfaces.}
\label{fig:s-disori}
\end{figure}

S-Al disorientations for S-phase precipitates situated at T-phase interfaces and non-interfacial counterparts are shown in \textbf{Fig. \ref{fig:s-disori}}, with previously reported S-Al ORs (\textbf{Table \ref{tab:SAl_ORs}}) highlighted. In both cases, the disorientations cluster together near the [100]$_{\text{S}}$//[001]$_\text{Al}$ axis. 
The cluster of disorientations for S-phases situated away from T-phase interfaces (\textbf{Fig. \ref{fig:s-disori}}(b)) is more clearly placed at the [100]$_{\text{S}}$//[001]$_\text{Al}$ axis, whereas the disorientations for interfacial S-phases (\textbf{Fig. \ref{fig:s-disori}}(a)) show a dense population slightly off axis. 
This implies that there exists more deviation from exact [100]$_{\text{S}}$//[001]$_\text{Al}$ parallelism when S-phase precipitates decorate the T-Al interfaces. S-phase precipitates placed away from T-Al interfaces are more strictly confined to in-plane ((001)$_\text{Al}$) rotations about [100]$_{\text{S}}$//[001]$_\text{Al}$ configuration.

S-Al interfaces for S-phase precipitates not decorating T-phases show a disorientation angle distribution shifted towards OR(I), with a dense population between OR(I) and OR(II) ($\sim$5$^{\circ}$ spread). For S-phases at T-phase interfaces the disorientations are more evenly spread across a $\sim$9$^{\circ}$ range of rotations. A 2$^{\circ}$ radius sphere of disorientations around OR(I-IV) accounts for 48\%, 35\%, 9\%, and 4\% of all disorientation data points in the case of S-phases at T-phase interfaces, respectively. Corresponding numbers are  78\%, 9\%, 3\%, and 1\% for S-phases located away from T-phase interfaces.

\subsubsection{S-T orientation relationships}

\begin{figure}[h]
\centering\includegraphics[width=0.85\linewidth]{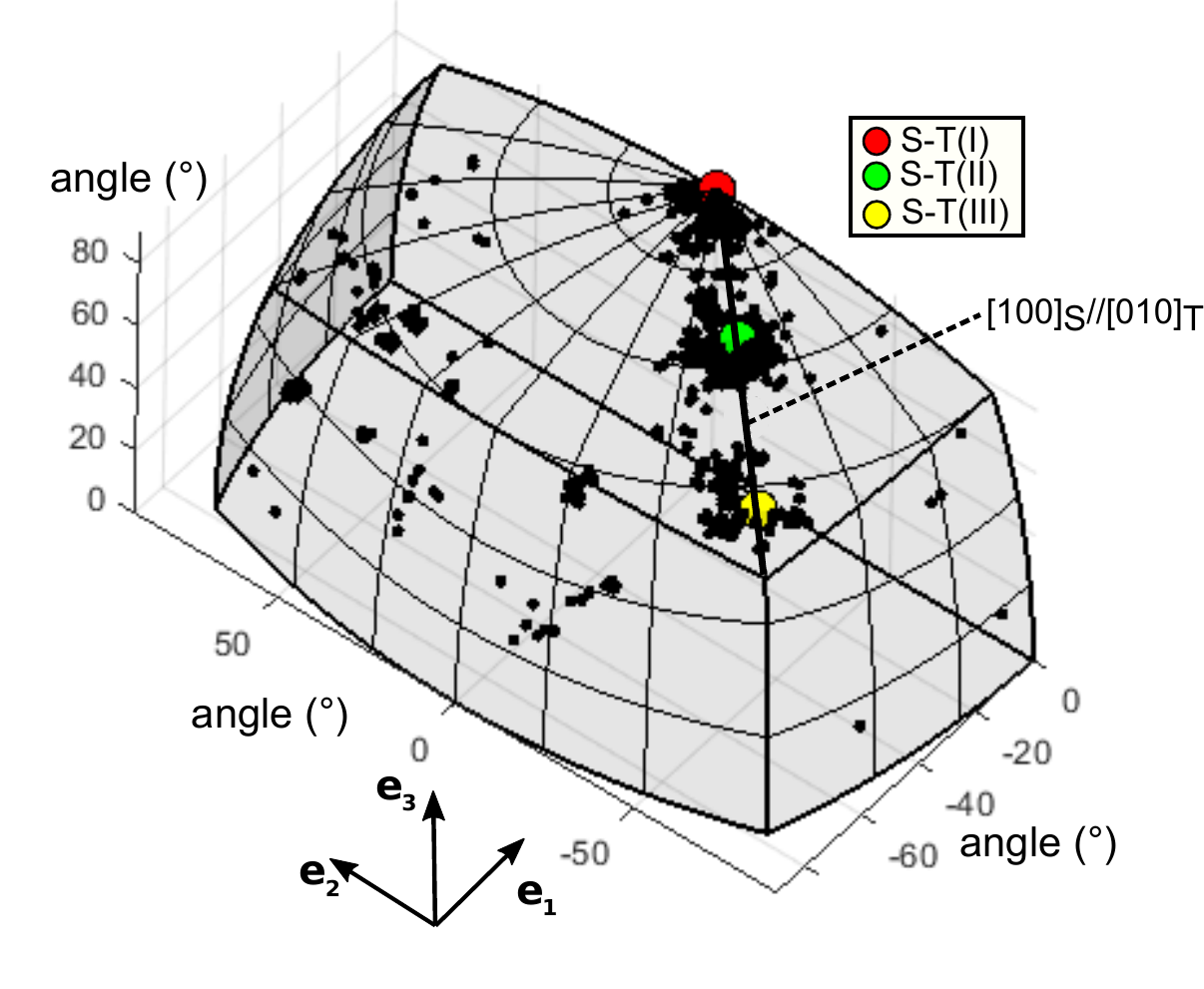}
\caption{S-T disorientations across 43 boundaries and proposed orientation relationships plotted in the corresponding fundamental zone of axis-angle space.}
\label{fig:s-t-disori}
\end{figure}

S-T phase boundary disorientations are shown in \textbf{Fig. \ref{fig:s-t-disori}}. The majority of disorientations fall close to the [100]$_{\text{S}}$//[010]$_{\text{T}}$ axis, and is consistent with HRTEM (\textbf{Fig. \ref{fig:mapping}}(a,d)) and HAADF-STEM (\textbf{Fig. \ref{fig:hrstem}}(d)) images, showing that the \textbf{a}- and \textbf{b}-axis of the S- and T-phase respectively, runs near parallel to [001]$_\text{Al}$. Several clear disorientation clusters can be seen, indicating the formation of several definite S-T crystallographic ORs. By sampling potential low index ORs placed along the [100]$_{\text{S}}$//[010]$_{\text{T}}$ axis, corresponding to a range of 90-120$^{\circ}$ in disorientation angles, the most probable crystallographic ORs were determined. The ORs were determined as those candidate ORs which accounted for the largest percentage of disorientations, falling within a $n$[$^{\circ}$] radius sphere of disorientations centred on the exact OR value, $n$ being an integer. 
The analysis yielded 3 ORs highlighted in \textbf{Fig. \ref{fig:s-t-disori}}. These ORs encompassed 26\%, 27\%, and 9\% of all data points, using a 3$^{\circ}$ radius sphere for ORs (I-III), respectively. The corresponding parallelisms and axis-angle representations are shown in \textbf{Table \ref{tab:ST_ORs}}.

\begin{table}[h]
\centering
\caption {S-T orientation relationships inferred from measured disorientations.}
\label{tab:ST_ORs}
\begin{tabular}{ c | c | c }
OR & Parallelism & Axis-angle \\
\hline
(I) & $\lbrace001\rbrace_{\text{S}}$ // $\lbrace001\rbrace_{\text{T}}$, $\langle100\rangle_{\text{S}}$ // $\langle010\rangle_{\text{T}}$ (// $\mathbf{n}$) & $\mathbf{n}, 90^{\circ}$ \\
(II) & $\lbrace011\rbrace_{\text{S}}$ // $\lbrace001\rbrace_{\text{T}}$, $\langle100\rangle_{\text{S}}$ // $\langle010\rangle_{\text{T}}$ & $\mathbf{n}, 96^{\circ}$ \\
(III) & $\lbrace013\rbrace_{\text{S}}$ // $\lbrace100\rbrace_{\text{T}}$, $\langle100\rangle_{\text{S}}$ // $\langle010\rangle_{\text{T}}$ & $\mathbf{n}, 112^{\circ}$
\end{tabular}
\end{table}

\subsection{Orientation spread}

\begin{figure*}[t]
\centering\includegraphics[width=.8\linewidth]{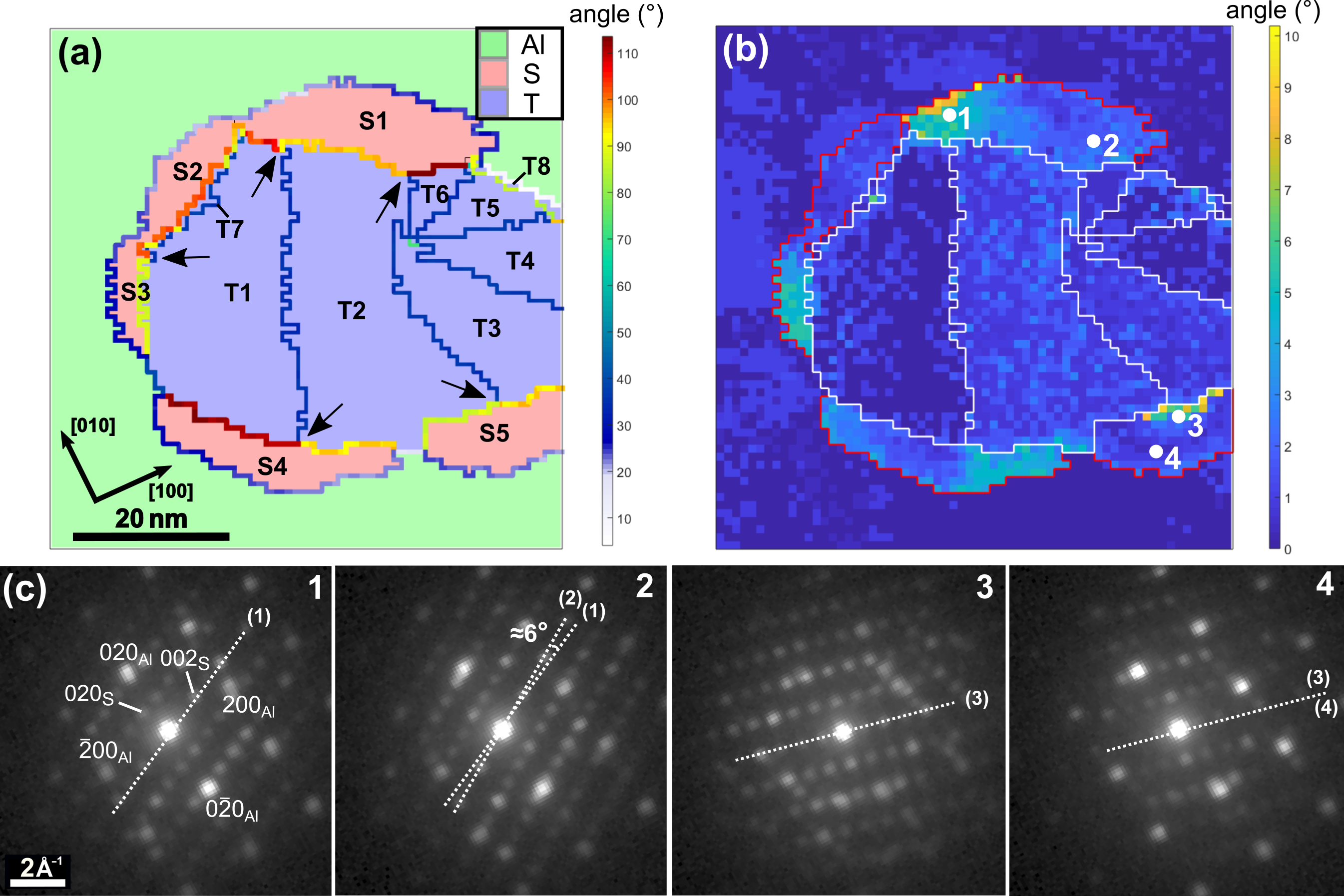}
\caption{(a) Phase map of a T-/S-phase aggregate with phase/domain boundaries coloured according to the misorientation angle between neighbouring pixels. Arrows highlight abrupt changes in misorientation angle across S-T boundaries. This is the same dispersoid aggragate as shown in \textbf{Fig. \ref{fig:mapping}}(a). (b) Misorientation with respect to the mean for each phase/domain, respectively. T-phase domains are highlighted by white lines, and red lines mark S-phase cross-sections. (c) The PED patterns from the numbered pixels 1-4 in (b). Dashed lines (1-4) from the associated PED pattern 1-4 is drawn to compare orientations.}
\label{fig:Crystallography2}
\end{figure*}

Disorientations between phases in the T-/S-phase aggregates generally showed significant spread about ORs described by plane parallelisms. This was investigated by direct inspection of SPED raw data and mapping results for each dispersoid aggregate, e.g. as shown in \textbf{Fig. \ref{fig:Crystallography2}}. The misorientation angle across S-T boundaries (\textbf{Fig. \ref{fig:Crystallography2}}(a)) varied over the range \SIrange{86}{113}{\degree}. The range was not continuously populated, but rather showed clustering about the angles corresponding with S-T ORs(I-III) (\textbf{Table \ref{tab:ST_ORs}}), consistent with the axis-angle representation of the combined S-T disorientations in \textbf{Fig. \ref{fig:s-t-disori}}. Spatially, the misorientation angle changed abruptly where the S-phase crossed T-phase TBs (indicated by arrows in \textbf{Fig. \ref{fig:Crystallography2}}(a)). The misorientation angle across S-Al boundaries spanned the range \SIrange{18}{27}{\degree}, again consistent with the combined axis-angle representation of S-Al disorientations for interfacial S-phase precipitates in \textbf{Fig. \ref{fig:s-disori}}(a). The population of angles showed a continuous, or near continuous distribution across this range. 
Particularly for S-phase precipitates crossing T-phase TBs, the full range of rotations was observed.

Taking S1 in \textbf{Fig. \ref{fig:Crystallography2}}(a) as an example, at the S1-T1 boundary the misorientation angles indicate that S-T OR(III) is followed. Moving from left to right crossing the T1-T2 TB, S1 makes a sharp adjustment at the S1-T2 interface to S-T OR(II) in order to accommodate the large changes in interface structure necessary to adjust to the rotation-twinned T2 domain. Crossing the T2-T6 TB, S-T OR(III) is re-established as the T-phase has rotated in the opposite direction back to T1 orientation (see \textbf{Fig. \ref{fig:mapping}}(c)). The transition involves $\sim$16$^{\circ}$ rotations of the S1-T interface. Adjustments are also observed at the phase boundary between S1 and the surrounding Al matrix. The S1-Al boundary exhibits a \SIrange{18}{27}{\degree} range of misorientation angles, gradually increasing from left to right. Comparing with reported plane parallelisms this nearly corresponds to a continuous or near continuous rotation from S-Al OR(I) to OR(IV). 

The misorientation with respect to the mean of each crystallographically distinct domain is shown in \textbf{Fig. \ref{fig:Crystallography2}}(b) and reveals significant variation inside individual S-phase precipitates. The largest deviations are observed where the S-phase crosses T-phase TBs. The deviation from the mean reach $\sim$9$^{\circ}$ for the interfacial S-phases, compared with variations below $\sim$3$^{\circ}$ within T-phase domains and the surrounding Al matrix.

The PED patterns presented in \textbf{Fig. \ref{fig:Crystallography2}}(c) are from marked positions in \textbf{Fig. \ref{fig:Crystallography2}}(b) and verify the discussed orientation variation of the T-/S-phase aggregates. The patterns show a significant in-plane ((001)$_{\text{Al}}$) rotation about the [100]$_{\text{S}}$//[001]$_{\text{Al}}$ axis within S1 (pattern 1-2), and an out-of-plane tilt of the S-phase structure away from this axis within S5 (pattern 3-4), which is apparent from the asymmetric intensity distribution and is consistent with the observation of phase boundary misorientations tilted away from simple parallelisms in addition to significant rotations about the parallel direction. As observed in axis-angle space, the largest tilts (spread of data points perpendicular to main axes) occur for S-phase precipitates in dispersoid aggregates (\textbf{Fig. \ref{fig:s-disori}}(a)), an example of which is shown here. 

\section{Discussion}

The combined TEM and SPED study presented here provides new insights to the inter-phase relations of T-/S-phase aggregates. In the following, these findings are interpreted in light of previous reports on the phases studied separately. 

The T-phase dispersoid forms during the homogenisation process at temperatures above $\SI{400}{\degreeCelsius}$ \cite{Zupanic}. Depending on the time and temperature at this stage the T-phase undergoes different degrees of rotation-twinning, which consequently affect the morphology. 
Long times and higher temperatures favour shell-shaped structures exhibiting pronounced twinning. A pseudo 10-fold symmetry centre can develop from successive twinning of T-twin domains. This symmetry centre is built from five differently oriented T-twin domains, each with a mirrored version, and which has three underlying T-Al ORs (\textbf{Table \ref{tab:TAl_ORs}}).  
A recent study \cite{Chen_new} has shown that the reported ORs can continually change as the T-phase undergoes rotation-twinning, requiring as much as 9$^{\circ}$ rotation of T-twin domains about the $[010]_\text{T}$//$[001]_\text{Al}$ axis, which likely results from the increase in interface energy caused by the twinning. 
The lattice mismatch of the T-phase hexagon subunit with (200)$_\text{Al}$ and (002)$_\text{Al}$ is largest for T-Al OR(III) (6.7\% and 8.6\%, respectively), which is only found in T-phases with multiple T-twin domains \cite{Chen3, Chen_new, Chen2}. Large rotations at the T-Al interface were also observed in the misorientation analysis conducted here, which measured up to $\sim$8$^{\circ}$ rotations about previously reported ORs (I-III), seen from \textbf{Fig. \ref{fig:t-disori}}(b). OR(III) was found to be the most frequent, which is reasonable as the majority of T-phase dispersoids analysed exhibited pronounced rotation-twinning (see \textbf{Fig. S1-20}).

Similarly, the misorientation analysis of non-interfacial S-phase precipitates showed several degrees variation from reported OR parallelisms. This was found to be approximately confined to a $\sim$5$^{\circ}$ range of rotations (\textbf{Fig. \ref{fig:s-disori}}(b)), a smaller number of data points reaching larger rotations. Previous reports show similar results, with upper bounds reaching $\SI{6}-\SI{9}{\degree}$ \cite{Winkelman, Kovarik, Radmilovic} rotation away from OR(I) reported by Bagaryatsky \cite{Bagaryatsky}. 
This attests to the preference for heterogeneous nucleation. 

It has hence been shown that both the S- and T-phase individually show in-plane ((001)$_\text{Al}$) rotations of several degrees from reported ORs with the Al matrix. The upper bounds of the range of rotation angles exhibited by S-phase precipitates situated away from T-phase interfaces corresponds well with the noted rotation that occur at the T-Al interface. The similarity may indicate why the T-Al interface is energetically favourable for S-phase nucleation and subsequent growth, and it reasons the finding of crystallographic ORs between the two phases. The correspondence of angles implies that the S-phase has the capability of accommodating the full rotation imposed by the maximum rotation at the T-Al interface. In the present study, S-phase precipitates were observed on all studied T-phase dispersoids, which suggests a good structural match between the two phases. These S-phases were seen to exhibit up to $\sim$9$^{\circ}$ rotation about an axis near $[100]_\text{S}$//$[001]_\text{Al}$ configuration (\textbf{Fig. \ref{fig:s-disori}}(a)), in close agreement with the magnitude of the range of rotations ($\sim$8$^{\circ}$) observed at the T-Al interface (\textbf{Fig. \ref{fig:t-disori}}(b)). As previously mentioned, the population of orientation angles exhibited by S-phase precipitates decorating T-phase interfaces was more spread towards larger angles than for the non-interfacial S-phases. This is to be expected as the S-phase has to accommodate the large rotations occurring at the T-Al interface.

The majority of S-phase precipitates in dispersoid aggregates seems to have grown preferentially along the T-Al interface. This is indicated by the observation that many S-phase precipitates in projection form caps at the prior T-Al interface, seen from \textbf{Fig. \ref{fig:mapping}}, \textbf{Fig. S1-20} and \textbf{Fig. S23-24}. This hence maximises the S-T interface area. Interfacial S-phase cross-sections typically have a narrow extension normal to the S-T interface. 
This implies that the S-T interface is energetically preferable as compared to S-Al interfaces. 
The interfacial S-phases have to adjust to local changes in both the T-phase structure and the surrounding Al matrix. The interface of the T-phase to which the S-phase has to conform is a complex rotation-twinned structure, often complicated further by regions of geometrical defects (see \textbf{Fig. \ref{fig:hrstem}} and \textbf{Fig. S23-24}). 
If the S-T interface incorporates a TB, significant adjustments of the S-phase structure must have resulted to maintain a definite crystallographic OR to each of the T-twin domains and the Al matrix, simultaneously. Furthermore, because the S-phase cross-section usually is narrow normal to the S-T interface, the structural adjustment must be a rather sharp transition in orientation in close proximity of the TB at the S-T interface. Misorientation angles across phase boundaries show that this is indeed the case, with the S-T interface shifting to a different OR across the TB (\textbf{Fig. \ref{fig:Crystallography2}}(a)). These changes in configuration involve rotations of several degrees of the S- and T-phase structures.  
The rotation is mainly about the [100]$_\text{S}$//[010]$_\text{T}$ axis, but tilts away from this configuration also occur (see \textbf{Fig. \ref{fig:s-t-disori}}).

The T-phase dispersoid often shows less orientation variation compared to the S-phase precipitates near the S-T phase boundaries, as the misorientation-to-mean-plot shows (\textbf{Fig. \ref{fig:Crystallography2}}(b)). This is because the phases are formed at different stages in material heat treatment. 
The rotations observed at T-Al interfaces must already be in place after quenching from homogenisation, unless the S-phase nucleation and growth can cause the T-phase dispersoid to bend. Due to the comparatively small size of interfacial S-phases relative to the T-phase dispersoid, this is not thought to occur. The S-phase nucleates and grows under artificial ageing ($\SI{170}{\degreeCelsius}$), which is too low in temperature for any significant changes in the T-phase structure to occur \cite{Zupanic, Chen_l}. The large local rotation of the S-phase structure near the TB is likely associated with a significant shear strain.

Some of the observed S-phase cross-sections at T-phase interfaces were likely formed by the coalescence of two initially separate S-phases. An example of this case was observed near a T-phase TB (see \textbf{Fig. S13}). The final morphology of S-phase precipitates in dispersoid aggregates can be seen as a balance between S-Al, S-T, and potentially S-S interface energies, reduction of lattice parameter misfits, and resulting shear strain.

The comparison between S-phase precipitates in dispersoid aggregates and S-phase precipitates formed away from T-phase interfaces provides further insight to the phase boundary energetics and the different inter-phase relations. Non-interfacial S-phases do not have to maintain a simultaneous OR with the T-phase in addition to the surrounding Al matrix. There is hence less need for large local adjustments in interface configurations, and the structure rotation that potentially arise from a variation in S- and/or (locally) Al lattice parameters \cite{Wang3, Winkelman, Kovarik} stays more confined about the [100]$_{\text{S}}$//[001]$_\text{Al}$ axis, as observed from \textbf{Fig. \ref{fig:s-disori}}(b). Furthermore, the range of orientation angles exhibited is smaller for non-interfacial S-phases ($\sim$5$^{\circ}$). Although S-phases at T-phase interfaces exhibit a broader range of disorientation angles with respect to the surrounding Al matrix ($\sim$9$^{\circ}$), the range is still approximately confined to the crystallographic limits proposed by Winkelman \textit{et al.} \cite{Winkelman} studying S-phases formed away from T-phase interfaces (\textbf{Fig. \ref{fig:s-disori}}(a)).

It is noteworthy that, the $\Omega$-phase, which was the most abundant phase in the studied alloy microstructure, exhibits a high resistance to coarsening \cite{Hutchinson} and remains small and uniformly dispersed. These precipitates are therefore likely the most significant contributors to strengthening in this over-aged condition. No $\Omega$-phase precipitates were observed at T-Al interfaces, which may be explained by S(S$'$) nucleating more easily and depleting the matrix of the solute supersaturation (mainly Cu) required to support the nucleation and growth of $\Omega$ \cite{Wang2}. 

\section{Conclusions}

SPED based orientation mapping combined with misorientation analysis in 3-dimensional axis-angle space and correlated HRTEM has been applied to study T-/S-phase aggregates in an Al-Cu-Mg-Ag alloy. The analysis revealed that:

\begin{itemize}
\item T-phase dispersoids show a rotation-twinned substructure characterised by $\sim$$\SI{36}{\degree}$ rotation about [010]$_{\text{T}}$ with \{101\}$_{\text{T}}$ as twin boundary planes. Dispersoids with a low number of twin domains ($<$3) tend to be lath-shaped, and T-phases showing more pronounced twinning have up to 10 domains, and exhibit shell-shaped cross-sections. 
\item T-Al interfaces follow the previously reported T-Al ORs(I-III), but show significant rotations ($\pm4^{\circ}$) about the [010]$_{\text{T}}$//[001]$_\text{Al}$ axis relative to exact OR parallelisms. 
\item S-Al interfaces for S-phases formed away from dispersoids show a disorientation angle distribution clustered about S-Al OR(I), mainly confined within a continuous (or near continuous) spread of $\sim$$5^{\circ}$ rotation towards OR(II). These S-phase precipitates showed a relatively strong confinement to rotations about the [100]$_{\text{S}}$//[001]$_\text{Al}$ axis.
\item S-Al interfaces for S-phase precipitates in dispersoid aggregates exhibit a continuous (or near continuous) range of disorientation angles between \SIrange{18}{27}{\degree} about the [100]$_{\text{S}}$//[001]$_\text{Al}$ axis. These rotations are roughly confined within the crystallographic limits between S-Al OR(I) and OR(IV) ($\sim$$9^{\circ}$  rotation).
In addition, these S-phases showed more out-of-plane tilts away from  [100]$_{\text{S}}$//[001]$_\text{Al}$ configuration compared to the non-interfacial counterparts.
\item Our work has found the following 3 ORs between S-phase precipitates and T-phase dispersoids:\\
(I)   $\lbrace001\rbrace_{\text{S}}$ // $\lbrace001\rbrace_{\text{T}}$, $\langle100\rangle_{\text{S}}$ // $\langle010\rangle_{\text{T}}$\\
(II)  $\lbrace011\rbrace_{\text{S}}$ // $\lbrace001\rbrace_{\text{T}}$, $\langle100\rangle_{\text{S}}$ // $\langle010\rangle_{\text{T}}$\\
(III) $\lbrace013\rbrace_{\text{S}}$ // $\lbrace100\rbrace_{\text{T}}$, $\langle100\rangle_{\text{S}}$ // $\langle010\rangle_{\text{T}}$
\item Axes-angle representation of S-Al, S-T and T-Al disorientations show several degrees deviation from reported/obtained crystallographic ORs. This is seen as a necessary consequence of changes in interface orientation and structure in maintaining simultaneous crystallographic ORs between S, T and Al.
\end{itemize}

We have demonstrated a methodology of correlating imaging and scanning diffraction techniques to allow precise analysis of orientation relationships and disorientations at high spatial resolution. The approach has potential for broad applications within multi-phase materials such as metals, semiconductors and minerals, where phase coherency is decisive for macroscopic properties.

\section*{Acknowledgements}

JKS and RH acknowledge support from the AMPERE project (NFR247783), a Knowledge-building Project for \mbox{Industry}, co-financed by The Research Council of Norway (NFR), and the industrial partners Hydro, Gr\"anges, \mbox{Neuman} Aluminium Raufoss (Raufoss Technology) and \mbox{Nexans}. RH and SW acknowledge funding from the Research Council of Norway NFR221714. PAM acknowledges funding from the ERC 291522-3DIMAGE, and the EPSRC Grant no. EP/R008779/1. DNJ acknowledges the EPSRC NanoDTC Cambridge EP/L015978/1 and the University of Cambridge for funding. The (S)TEM work was carried out on the NORTEM \mbox{infrastructure} (NFR197405) at the TEM Gemini Centre, Trondheim, \mbox{Norway}.

\section*{References}

\end{document}